\documentclass[12pt]{article}
\usepackage{amsmath}
\usepackage{amsfonts}
\usepackage{amssymb}
\usepackage{graphicx}

\input{tcilatex}
\begin{document}

\begin{titlepage}
%\begin{flushright}
%2008/3/2
%\end{flushright}
\ \\
\begin{center}
\LARGE
{\bf
Quantum Measurement Information\\
as a key to Energy Extraction\\
 from Local Vacuums\\
}
\end{center}
\ \\
\begin{center}
\large{
Masahiro Hotta
}\\
\ \\
\ \\
{\it
Department of Physics, Faculty of Science, Tohoku University,\\
Sendai, 980-8578, Japan\\
hotta@tuhep.phys.tohoku.ac.jp
}
\end{center}
\begin{abstract}
In this paper, a protocol is proposed in which energy extraction from local vacuum states is possible by using  
quantum measurement information for the vacuum state of quantum fields.   
In the protocol,  Alice, who stays at a spatial point, excites the ground state of the fields by a local measurement. 
 Consequently, wavepackects generated by A' measurement  propagate the vacuum to spatial infinity. Let us assume that   
Bob  stays away from Alice and fails to catch the excitation energy when the wavepackets pass in front of him.   
 Next Alice announces her local measurement result to Bob by classical communication. 
Bob performs a local unitary operation depending on the measurement result. In this process, positive energy
 is released  from the fields to Bob's apparatus of the unitary operation. In the field systems, wavepackets are
 generated with negative energy around Bob's location. Soon afterwards, the negative -energy wavepackets begin to chase 
after the positive-energy wavepackets generated by Alice and form loosely bound states.
\end{abstract}

\end{titlepage}

\bigskip

\section{Introduction}

\ \newline

The interplay between quantum information theory and quantum field theory
has intensified and is expected to revolutionize physics. For example, novel
ideas are proposed on, for example, the information loss problem of black
holes\cite{qbh} and quantum causal histories of quantum gravity\cite{qcs}.
It has been also discussed that distillation of vacuum-entanglement of
quantum fields yields EPR pairs \cite{R1} and W states \cite{R2}. In
reference \cite{hotta}, the method of field theory in curved spacetime is
employed to evaluate actuating energy of photon switching in quantum
communication. In this paper, we apply positive operator value measure
(POVM) and local operations and classical communication (LOCC) to the
physics of negative energy density in quantum field theory. POVM and LOCC
are fundamental tools of quantum information theory\cite{text}.

Quantum fluctuations of the local energy around its zero value in field
theory has been studied for a long time \cite{BD}. Quantum interference is
able to create states containing regions of negative energy, though the
total energy remains nonnegative. A notion of negative energy has impacts on
many fundamental problems of physics, including traversable wormhole \cite%
{wt}, cosmic censorship \cite{cc} and the second law of thermodynamics \cite%
{2nd}. \ It has been pointed out that available absolute values of negative
energy are crucial for those problems. Possible values of negative energy
are restricted by quantum inequalities for energy density based on
uncertainty relations \cite{qi}\cite{F}.

Classical energy of free fields takes nonnegative values, and cannot be used
successfully when our apparatus of energy extraction from the fields is
located outside nonvanishing-energy regions. This situation is dramatically
different for quantum energy. Let us consider a local quantum measurement
performed for the vacuum state. A finite amount of positive energy is
infused into the system at the measurement device position. Because the
properties of states excited by local measurement are the same as those of
the vacuum outside the excited regions, those states can be called local
vacuum states. The concept of local vacuum states is the same of that of
strictly localized states proposed by Knight \cite{Knight}. In this paper, \
it is proven for a free massless scalar field in 1+1 dimensions that the
excitation energy can be partly extracted back from the field using the
measurement results and a quantum apparatus located away from the
measurement point, \textit{even if the field has, on average, no energy
around the apparatus at all}. Using this method, we can transport energy to
a distant location by sending not a physical object with excitation energy,
but classical information. In the extraction process, wavepackets with
negative energy density are generated in the system \ and form loosely bound
states with positive-energy wavepackets excited first by the measurement
device. This method is based on a quantum energy teleportation protocol
proposed for spin chains \cite{hotta2}. \ The protocol transfers localized
energy from one site of a spin chain to another only by LOCC. However, aims
of the paper are confined to short-time-scale processes in which dynamical
evolution induced by the Hamiltonian is negligible, although LOCC is assumed
possible many times in the short interval. In relativistic field systems,
the dynamical effect propagates with light velocity, which is the upper
bound on the speed of classical communication. Thus, we generally cannot
omit global time evolution. It is also noted that any continuous limit of
zero lattice spacing cannot be taken for the protocol in \cite{hotta2}
because measurements in the protocol are projective, which becomes an
obstacle to obtaining a smooth limit. In this paper, we adopt a different
general measurement that is well defined in field theory. \ 

\bigskip

The paper is organized as follows. In section 2, we briefly review general
measurements and LOCC of quantum information\ theory. In section 3,\ a short
review of negative-energy physics of \ a 1+1 dimensional free scalar field
is given. Section 4 presents a protocol in which energy is extracted from
local vacuum states using measurement results. In section 5, an explicit
example of the protocol is given. Section 6 summarizes the results. We adopt
the natural unit\bigskip $c=\hbar =1$.

\section{POVM and LOCC}

\bigskip

In this section, we give an overview of concepts related to general
measurements by use of POVM and measurement operator and LOCC in quantum
information theory. A detailed explanation can be found in standard
textbooks of quantum information\cite{text}. measurements are generalized
measurements beyond projective (ideal) measurements. Let us consider a
quantum system $S$ about which we wish to obtain information. In order to
formulate general measurements, we need another quantum system $S^{\prime }$
as a probe. In general, dimension $N$ of the Hilbert space of $S$ is not
equal to that of $S^{\prime }$. We bring $S^{\prime }$ into contact with $S$
by probe interactions between the two. In this process, information on $S$
is imprinted into $S^{\prime }$. After switch-off of the measurement
interactions, we perform a projective measurement on not $S$ but the probe
system $S^{\prime }$ and obtain imprinted information about $S$. This
completes a general measurement. An ideal measurement can be made if a
composite quantum state after switch-off of the interaction is given by

\begin{equation*}
|\Psi \rangle _{SS^{\prime }}=\sum_{n=1}^{N}c_{n}|n\rangle _{S}|u_{n}\rangle
_{S^{\prime }},
\end{equation*}%
where $\left\{ |n\rangle |n=1\sim N\right\} $ is the complete set of
orthonormal basis state vectors of $S$ and $\left\{ |u_{n}\rangle
_{S^{\prime }}\right\} $ is the set of orthonormal state vectors of $%
S^{\prime }$. When a measurement result for $S^{\prime }$ is given by $%
|u_{n}\rangle _{S^{\prime }}$ with probability $p_{n}=|c_{n}|^{2}$, we infer
that $S$ is also observed in the state $|n\rangle _{S}$ with the same
probability. Hereafter we will express quantum states by density operators.
General measurements are mathematically described using measurement
operators $M_{\mu }$ ($\mu =1\sim m$), which act on the Hilbert space of $S$
and satisfy

\begin{equation}
\sum_{\mu =1}^{m}M_{\mu }^{\dag }M_{\mu }=I_{S},  \label{1}
\end{equation}%
where the number of $M_{\mu }$ is denoted by $m$ and generally not equal to $%
N$. Let us consider explicitly an indirect measurement model in order to
understand the measurement operators. Let us write down a probe Hamiltonian
as 
\begin{equation*}
H_{p}(t)=\sum_{\gamma }g_{\gamma }(t)O_{S}^{(\gamma )}\otimes O_{S^{\prime
}}^{(\gamma )},
\end{equation*}%
where $O_{S}^{(\gamma )},O_{S^{\prime }}^{(\gamma )}$ are Hermitian
operators acting on the Hilbert spaces of $\,S$ and $S^{\prime }$, and $%
g_{\gamma }(t)$ are real functions of time $t$ which take zero values for $%
t\notin \left( 0,T\right) $. The interaction generates entanglement between $%
S$ and $S^{\prime }$. The time evolution is described by the following
unitary operator.

\begin{equation*}
U_{p}(T)=\limfunc{T}\exp \left[ -i\int_{0}^{T}H_{p}(t)dt\right] =\exp \left[
-i\sum_{\gamma }\int_{0}^{T}g_{\gamma }(t)dtO_{S}^{(\gamma )}\otimes
O_{S^{\prime }}^{(\gamma )}\right] .
\end{equation*}%
Let us set the initial state as $|\psi _{S}\rangle $ for $S$, and $%
|0_{S^{\prime }}\rangle $ for $S^{\prime }$ at $t=0$. After switch off of
the probe interaction, the total state is given by

\begin{equation*}
|\Phi \rangle =U_{p}(T)\left( |\psi _{S}\rangle \otimes |0_{S^{\prime
}}\rangle \right) .
\end{equation*}%
Now let us perform a $S^{\prime }~$projective measurement for $|\Phi \rangle 
$. Consider a complete orthonormal basis $\left\{ |\mu ,S^{\prime }\rangle
|\mu =1,\cdots ,m\right\} $of the Hilbert space of $S^{\prime }$. The index $%
\mu $ classifies $m$ possible outputs of the measurement. The projection
operator onto $|\mu ,S^{\prime }\rangle $ is defined by

\begin{equation*}
P_{\mu }(S^{\prime })=|\mu ,S^{\prime }\rangle \langle \mu ,S^{\prime }|.
\end{equation*}%
Because of completeness, the following relation is satisfied.

\begin{equation}
\sum_{\mu =1}^{m}P_{\mu }(S^{\prime })=I_{S^{\prime }}.  \label{pj}
\end{equation}%
The measurement operator $M_{\mu }$ is obtained by acting $I_{S}\otimes
P_{\mu }(S^{\prime })$ on $|\Phi \rangle $ such that

\begin{equation*}
\left( I_{S}\otimes P_{\mu }(S^{\prime })\right) |\Phi \rangle =M_{\mu
}|\psi _{S}\rangle \otimes |\mu ,S^{\prime }\rangle .
\end{equation*}%
It is noted that $M_{\mu }$ are operators acting on the Hilbert space of $S$%
. The explicit form of $M_{\mu }$ is given by

\begin{equation*}
M_{\mu }=\langle \mu ,S^{\prime }|U_{p}(T)|0_{S^{\prime }}\rangle .
\end{equation*}
Eq.(\ref{1}) \ then is easily verified as follows.

\begin{eqnarray*}
\sum_{\mu =1}^{m}M_{\mu }^{\dag }M_{\mu } &=&\sum_{\mu =1}^{m}\langle
0_{S^{\prime }}|U_{p}^{\dag }(T)|\mu ,S^{\prime }\rangle \langle \mu
,S^{\prime }|U_{p}(T)|0_{S^{\prime }}\rangle \\
&=&\sum_{\mu =1}^{m}\langle 0_{S^{\prime }}|U_{p}^{\dag }(T)\left(
I_{S}\otimes |\mu ,S^{\prime }\rangle \langle \mu ,S^{\prime }|\right)
U_{p}(T)|0_{S^{\prime }}\rangle \\
&=&\langle 0_{S^{\prime }}|U_{p}^{\dag }(T)\left( I_{S}\otimes \sum_{\mu
=1}^{m}P_{\mu }(S^{\prime })\right) U_{p}(T)|0_{S^{\prime }}\rangle \\
&=&\langle 0_{S^{\prime }}|U_{p}^{\dag }(T)U_{p}(T)|0_{S^{\prime }}\rangle \\
&=&\langle 0_{S^{\prime }}|I_{S}\otimes I_{S^{\prime }}|0_{S^{\prime
}}\rangle =I_{S}.
\end{eqnarray*}%
In the above proof, we have used Eq.(\ref{pj}) and unitarity of $U_{p}(T)$.
It should be stressed that in general, $M_{\mu }$ is not a projective
operator. It can be shown \cite{text} that for an arbitrary quantum state $%
\rho $ of $S$, the result $\mu $ is observed with probability $p_{\mu }$
evaluated via

\begin{equation}
p_{\mu }=\limfunc{Tr}\left[ \rho M_{\mu }^{\dag }M_{\mu }\right] .  \label{2}
\end{equation}%
After the measurement, the state of $S$ is transformed into a state given by

\begin{equation}
\rho _{\mu }=\frac{M_{\mu }\rho M_{\mu }^{\dag }}{\limfunc{Tr}\left[ \rho
M_{\mu }^{\dag }M_{\mu }\right] }.  \label{3}
\end{equation}%
These results are always correct when we start from any indirect measurement
model. It has been proven \cite{ozawa2} that, inversely, if we have some
operators $M_{\mu }$ satisfying Eq.(\ref{1}), there exists an indirect
measurement model with a probe system $S^{\prime }$ and a measurement
interaction between $S$ and $S^{\prime }$ such that the relations in Eq.(\ref%
{2}) and Eq.(\ref{3}) are reproduced. Hence, we are able to make general
arguments on general measurements by considering general operators $M_{\mu }$
satisfying Eq.(\ref{1}). In mathematics, the set of Hermitian positive
semidefinite operators $M_{\mu }^{\dag }M_{\mu }$ \ is called a positive
operator value measure (POVM). This is because some people call the general
measurement POVM measurement.

Here I give comments for measurements in field theory. The localized general
measurement operators are expressed as functions of averaged local operators
with test functions with compact supports. For example, let us consider
arbitrary local operators $O_{k}(x)$ with $k=1,2,\cdots ,\infty $ of a field
system in one spatial dimension, $S$. Then the averaged operators are given
by%
\begin{equation*}
\bar{O}_{k}(R)=\int_{-\infty }^{\infty }\omega _{R}(x)O_{k}(x)dx,
\end{equation*}%
where $\omega _{R}(x)$ is a test function with a compact support $R$. The
Hamiltonian of probe interactions depends on time $t$ and those averaged
operators as follows. 
\begin{equation*}
H_{p}=H_{p}\left( t,\bar{O}_{1}(R_{1}),\bar{O}_{2}(R_{2}),\cdots \right) .
\end{equation*}%
The time evolution operator is given by%
\begin{equation*}
U_{p}(T)=\limfunc{T}\exp \left[ -i\int_{0}^{T}H_{p}\left( t,\bar{O}%
_{1}(R_{1}),\bar{O}_{2}(R_{2}),\cdots \right) dt\right] .
\end{equation*}%
The general measurements for fields are fixed by giving $H_{p}~$and the
initial state $|S^{\prime }\rangle $ of the probe system $S^{\prime }$. The
final state of $S,$ after the ideal measurement of the probe system $%
S^{\prime }$ yields the result $\mu $, is expressed by use of the
measurement operators $M_{\mu }$ as follows.

\begin{equation}
\limfunc{Tr}_{S^{\prime }}\left[ P_{\mu }(S^{\prime })\left( U_{p}(T)\left(
|S\rangle \langle S|\otimes |S^{\prime }\rangle \langle S^{\prime }|\right)
U_{p}^{\dag }(T)\right) \right] =M_{\mu }|S\rangle \langle S|M_{\mu }^{\dag
},  \label{povm}
\end{equation}%
where $|S\rangle $ is an arbitrary initial state of $S$. It is then noticed
that $M_{\mu }$ becomes a function of the averaged operators as 
\begin{equation*}
M_{\mu }=M_{\mu }\left( \bar{O}_{1}(R_{1}),\bar{O}_{2}(R_{2}),\cdots \right)
.
\end{equation*}%
Even if one wants to take a non-separable initial state $|S+S^{\prime
}\rangle $ of $S$ and $S^{\prime }$ in Eq. (\ref{povm}), \ the formulation
discussed above still works. This is because any $|S+S^{\prime }\rangle $ is
reproduced by acting a unitary operation $V~$on a separable state $%
|S_{I}\rangle |S_{I}^{\prime }\rangle :$

\begin{equation*}
|S+S^{\prime }\rangle =V|S_{I}\rangle |S_{I}^{\prime }\rangle .
\end{equation*}%
Therefore we are able to introduce measurement operators $\tilde{M}_{\mu }$
satisfying 
\begin{eqnarray*}
&&\limfunc{Tr}_{S^{\prime }}\left[ P_{\mu }(S^{\prime })\left(
U_{p}(T)|S+S^{\prime }\rangle \langle S+S^{\prime }|U_{p}^{\dag }(T)\right) %
\right] \\
&=&\limfunc{Tr}_{S^{\prime }}\left[ P_{\mu }(S^{\prime })\left(
U_{p}(T)V\left( |S_{I}\rangle \langle S_{I}|\otimes |S_{I}^{\prime }\rangle
\langle S_{I}^{\prime }|\right) V^{\dag }U_{p}^{\dag }(T)\right) \right] \\
&=&\tilde{M}_{\mu }|S_{I}\rangle \langle S_{I}|\tilde{M}_{\mu }^{\dag }.
\end{eqnarray*}

The general measurements are generated by probing interactions (expressed by 
$H_{p}$) which are assumed to be switched on in a time interval, $\left[ 0,T%
\right] $. Effective switching of those couplings may be achieved by various
methods in field theory. For example, by applying laser beams to
semiconductor devices in quantum optics, energy levels of the devices can be
shifted corresponding to the beam strength. This mechanism has been applied
to control of photon-counter switching.

LOCC is a setting of quantum communication. Let us consider two parties who
share a quantum state of a composite system and want to communicate with
each other using the quantum system and classical channels. In the LOCC
setting, they are able to perform local operations at each side, including
local unitary transformations and local general measurements. The two
parties are also allowed to use classical channels for sending classical
information like measurement results. However, they are not allowed to use
global quantum operations over the composite system. For example, quantum
teleportation \cite{qt} is a well-known protocol obtainable by LOCC. It
transfers any unknown quantum state to a distant place.

\bigskip

\section{Negative Energy Density of Quantum Fields}

\bigskip

In this section, we give an overview of negative-energy physics of a 1+1
dimensional free scalar field $\phi $. The properties described will be
applied to a protocol in the next section. A detailed explanation can be
found in \cite{hotta} and \cite{F}. The equation of motion is%
\begin{equation}
\left[ \frac{\partial ^{2}}{\partial t^{2}}-\frac{\partial ^{2}}{\partial
x^{2}}\right] \phi (t,x)=0.  \label{eqm}
\end{equation}%
The general solution of Eq.\ (\ref{eqm}) is written as a sum of left- and
right-moving components: $\phi (x,t)=\phi _{+}\left( x^{+}\right) +\phi
_{-}\left( x^{-}\right) $, where $\phi _{+}\left( x^{+}\right) $ denotes the
left-moving field and $\phi _{-}\left( x^{-}\right) $ the right-moving field
with light-cone coordinates $x^{\pm }=t\pm x.$ It is remarkable that the
quantum interference effect between multi-particle states is able to
suppress quantum fluctuation of the field and to yield negative energy
density of the field. For example, even though the classical energy flux $%
\left[ \partial _{+}\phi _{+}(x^{+})\right] ^{2}$ of the left-moving field
is nonnegative, the expectation value of the corresponding quantum flux
operator $T_{++}\left( x^{+}\right) =:\partial _{+}\phi _{L}(x^{+})\partial
_{+}\phi _{L}(x^{+}):$ can be negative. Despite the existence of regions
with negative energy density, expectation values of the total energy flux $%
\int_{-\infty }^{\infty }T_{++}(x^{+})dx^{+}$ for an arbitrary state remain
nonnegative because the total flux is given by $\int_{0}^{\infty }\hbar
\omega a_{\omega }^{+\dag }a_{\omega }^{+}d\omega $. By taking an arbitrary,
monotonically increasing $C^{1}~$function $f(x)$of $x\in \left( -\infty
,\infty \right) $satisfying $f(\pm \infty )=\pm \infty $, the set of mode
functions

\begin{equation}
v_{\omega }(x)=\sqrt{\frac{\hbar }{4\pi \omega }}e^{-i\omega f(x)},~(\omega
\geq 0)  \label{vm}
\end{equation}%
is obtained, which can uniquely expand the field. Their orthonormality in
terms of the normal product can also be derived straightforwardly. By using
mode functions, the left-moving field $\phi _{+}$ is expanded via

\begin{equation*}
\phi _{+}\left( x^{+}\right) =\int_{0}^{\infty }d\omega \left[ b_{\omega
}^{+}v_{\omega }(x^{+})+b_{\omega }^{+\dag }v_{\omega }^{\ast }(x^{+})\right]
.
\end{equation*}%
Here $b_{\omega }^{+\dag },~b_{\omega }^{+}$ are creation and annihilation
operators that satisfy $\left[ b_{\omega }^{+},~b_{\omega ^{\prime }}^{+\dag
}\right] =\delta \left( \omega -\omega ^{\prime }\right) $. We note that the
normalized quantum state $|\Phi \rangle $ defined by $b_{\omega }^{+}|\Phi
\rangle =0$ is a squeezed state. For $|\Phi \rangle $, the expectation value
is evaluated through

\begin{equation}
\langle \Phi |T_{++}\left( x^{+}\right) |\Phi \rangle =-\frac{\hbar }{24\pi }%
\left[ \frac{\dddot{f}\left( x^{+}\right) }{\dot{f}\left( x^{+}\right) }-%
\frac{3}{2}\left( \frac{\ddot{f}\left( x^{+}\right) }{\dot{f}\left(
x^{+}\right) }\right) ^{2}\right] ,  \label{f}
\end{equation}%
where the dot denotes a derivative in terms of $x^{+}$\cite{BD}. \ An
interesting example of negative energy flux is generated by a monotonically
increasing $C^{1}$ function $f_{\varepsilon }\left( x\right) $ given by

\begin{align*}
f_{\varepsilon }\left( x\right) & =\Theta \left( x_{i}-x\right) x \\
& +\Theta \left( x_{f}-x\right) \Theta \left( x-x_{i}\right) \left[ x_{i}-%
\frac{1}{\sqrt{\varepsilon }}+\frac{1}{\sqrt{\varepsilon }-\varepsilon
\left( x-x_{i}\right) }\right] \\
& +\Theta \left( x-x_{f}\right) \left[ \frac{\varepsilon }{\left( \sqrt{%
\varepsilon }-\varepsilon \left( x_{f}-x_{i}\right) \right) ^{2}}%
(x-x_{f})+x_{i}-\frac{1}{\sqrt{\varepsilon }}+\frac{1}{\sqrt{\varepsilon }%
-\varepsilon \left( x_{f}-x_{i}\right) }\right] ,
\end{align*}%
where $x_{i}\leq x_{f}$, $\Theta \left( x\right) $ is a step function and $%
\varepsilon =\left( \frac{12\pi \left\vert E_{n}\right\vert }{\hbar }\right)
^{2}$ is a nonnegative constant. For the squeezed state $|\Phi
_{shock}\rangle $ \ corresponding to $f_{\varepsilon }\left( x\right) $, \
the left-moving energy flux is estimated by

\begin{equation}
\langle \Phi _{shock}|T_{++}(x^{+})|\Phi _{shock}\rangle =-\left\vert
E_{n}\right\vert \delta (x^{+}-x_{i})+\frac{\left\vert E_{n}\right\vert }{1-%
\frac{12\pi }{\hbar }\left\vert E_{n}\right\vert l}\delta \left(
x^{+}-x_{f}\right) ,  \label{n+p}
\end{equation}%
where $l=x_{f}-x_{i}(>0)$. The first term on the right-hand side shows the
flux of a shock wave with negative energy $-\left\vert E_{n}\right\vert $.
Because $\int_{-\infty }^{\infty }\langle \Phi _{shock}|T_{++}\left(
x\right) |\Phi _{shock}\rangle dx$ is positive, we obtain the following
inequality%
\begin{equation*}
\frac{\left\vert E_{n}\right\vert ^{2}l}{\frac{\hbar }{12\pi }-\left\vert
E_{n}\right\vert l}\geq 0.
\end{equation*}%
Because the numerator is definitely positive, the denominator must be
nonnegative, which leads to an uncertainty-relation-type inequality: 
\begin{equation}
l=x_{f}-x_{i}\leq \frac{\hbar }{12\pi \left\vert E_{n}\right\vert }.
\label{nieq}
\end{equation}%
This means that negative-energy shock waves cannot be separated infinitely
far from positive-energy shock waves. This is because the existence of
negative energy is sustained by a quantum correlation effect with
positive-energy excitations. If the quantum correlation vanishes completely,
negative energy cannot appear in any region because nonnegativity of the
Hamiltonian should hold in every local region. Hence, it can be concluded
that the negative-energy shock waves form loosely bounded states with the
positive-energy shockwaves. Creation of the above loosely bound states of
negative- and positive-energy excitations is not peculiar to this example,
but rather takes place in any arbitrary system with negative-energy local
excitations.

\section{Energy Extraction from Local Vacuums by LOCC}

\bigskip

\bigskip

In this section, a protocol for energy extraction from local vacuums by LOCC
is proposed for a free massless scalar field $\phi $ in 1+1 dimensions. The
system is introduced as a toy model to present a new idea which can be
applied to 3+1 dimensional electromagnetic field. We may consider that $\phi 
$ corresponds to asymptotic field of QED gauge field in the analogy.
Hereinafter, we will refer to this protocol as quantum field energy
teleportation (QFET). In canonical quantization, the standard commutation
relations are set for the canonical Schr\H{o}dinger operators as follows. 
\begin{equation*}
\left[ \hat{\phi}\left( x\right) ,~\hat{\Pi}\left( x^{\prime }\right) \right]
=i\delta \left( x-x^{\prime }\right) ,
\end{equation*}

\begin{equation*}
\left[ \hat{\phi}(x),~\hat{\phi}(x^{\prime })\right] =0,
\end{equation*}%
\begin{equation*}
\left[ \hat{\Pi}\left( x^{\prime }\right) ,~\hat{\Pi}\left( x^{\prime
}\right) \right] =0.
\end{equation*}%
The energy density operator is written as

\begin{equation*}
\hat{\varepsilon}(x)=\frac{1}{2}\left[ \hat{\Pi}^{2}+\left( \partial _{x}%
\hat{\phi}\right) ^{2}\right] -\varepsilon _{0},
\end{equation*}%
where $\varepsilon _{0}$ is a constant for subtraction of the vacuum
contribution. The Hamiltonian is given by spatial integration of $\hat{%
\varepsilon}(x)$ as $\hat{H}=\int \hat{\varepsilon}(x)dx$. The vacuum $%
|0\rangle $ is the eigenstate corresponding to the lowest eigenvalue of $%
\hat{H}$. By adjusting $\varepsilon _{0}$, we can set

\begin{equation*}
\langle 0|\hat{\varepsilon}\left( x\right) |0\rangle =0,
\end{equation*}%
\begin{equation*}
\hat{H}|0\rangle =0.
\end{equation*}%
This choice of $\varepsilon _{0}$ corresponds to the normal order
prescription. The evolution operator of the system is defined by $U(t)=e^{-it%
\hat{H}}$. Then, using the Schr\H{o}dinger operators, the canonical
Heisenberg operators are calculated as

\begin{align}
\hat{\phi}(t,x)& =\frac{1}{2}\left[ \hat{\phi}(x+t)+\hat{\phi}(x-t)\right] 
\notag \\
& +\frac{1}{2}\int_{x-t}^{x+t}\hat{\Pi}(y)dy,  \label{103}
\end{align}

\begin{align}
\hat{\Pi}(t,x)& =\frac{1}{2}\left[ \hat{\Pi}(x+t)+\hat{\Pi}(x-t)\right] 
\notag \\
& +\frac{1}{2}\left[ \partial _{x}\hat{\phi}(x+t)-\partial _{x}\hat{\phi}%
(x-t)\right] .  \label{104}
\end{align}

Let us consider Alice at $x=x_{A}~$who excites the ground state of the field
by a local measurement, and Bob who stays at $x=x_{B}$ away from Alice and
extracts energy from the field. Then the QFET protocol is composed of the
following four phases:

\bigskip

(1) At time $t=0$, Alice makes a local general measurement defined by
operators $M_{n}\left( A\right) $ satisfying

\begin{equation}
\sum_{n}M_{n}^{\dag }\left( A\right) M_{n}\left( A\right) =1  \label{5}
\end{equation}%
to the vacuum state $|0\rangle $ and obtains the result $n$. To perform this
measurement, she must, on average, give positive energy $E_{A}$ to the
field. Using this energy, positive-energy wavepackets of the field are
generated.

\bigskip\ 

(2) At time $t=t_{o}$, the wavepackets excited by Alice have already passed
by the position of Bob. Assume that Bob fails to catch any energy of the
wavepackets at all. Consequently, no energy of $\phi $ remains around Bob
after $t=t_{o}$.

\bigskip

(3) Alice announces the measurement result $n~$to Bob by classical
communication. Bob receives the\ information at time $t=T(\geq t_{o})$.

\bigskip

(4) At $t=T$, Bob performs a unitary operation depending on the value of $n$
defined by

\begin{equation}
U_{n}\left( B\right) =\exp \left[ iga_{n}\int_{-\infty }^{\infty }p_{B}(x)%
\hat{\phi}(x)dx\right] ,  \label{16}
\end{equation}%
where $g$ is a real constant fixed below, $a_{n}$ are real constants
depending on $n$ and $p_{B}(x)$ is a function whose support is localized
around Bob's location. In this process, positive energy $E_{B}$ is released
on average from $\phi $ to Bob's apparatus of $U_{n}\left( B\right) $. In
the system of $\phi $, wavepackets are generated with negative energy $%
-E_{B} $ around Bob's location. Soon afterwards, the wavepackets begin to
chase after the positive-energy wavepackets generated by Alice.

The schematics in Figures 1 to 3 describe this QFET protocol with plots of $%
\langle \varepsilon (x)\rangle =\limfunc{Tr}\left[ \rho \hat{\varepsilon}(x)%
\right] $ as a function of $x$. A spacetime diagram for protocol events is
given in Figure 4. The amount of energy $E_{A}$ is evaluated by%
\begin{equation*}
E_{A}=\sum_{n}\langle 0|M_{n}^{\dag }\left( A\right) \hat{H}M_{n}\left(
A\right) |0\rangle >0.
\end{equation*}%
After phase (1), the quantum state is transformed into the following state
depending on\bigskip\ $n$.%
\begin{equation}
|A_{n}\rangle =\frac{1}{\sqrt{\langle 0|M_{n}^{\dag }\left( A\right)
M_{n}\left( A\right) |0\rangle }}M_{n}\left( A\right) |0\rangle .  \label{10}
\end{equation}%
The average quantum state after measurement evolves until $t=T$ as follows:%
\begin{equation}
\rho (T)=\sum_{n}U(T)M_{n}\left( A\right) |0\rangle \langle 0|M_{n}^{\dag
}\left( A\right) U^{\dag }(T).  \label{7}
\end{equation}%
Soon after phase (4), the average quantum state transforms into the
following state:

\begin{equation*}
\rho _{F}=\sum_{n}U_{n}\left( B\right) U(T)M_{n}\left( A\right) |0\rangle
\langle 0|M_{n}^{\dag }\left( A\right) U^{\dag }(T)U_{n}^{\dag }\left(
B\right) .
\end{equation*}%
In order to evaluate $E_{B}$, let us introduce a localized energy operator $%
\phi $ around Bob:

\begin{equation*}
\hat{H}_{B}=\int w_{B}\left( x\right) \hat{\varepsilon}\left( x\right) dx,
\end{equation*}%
where $w_{B}$ is a nonnegative window function that satisfies

\begin{equation*}
w_{B}\left( x\right) =1
\end{equation*}%
for $x\in \left( x_{B}-\epsilon ,x_{B}+\epsilon \right) $ with a positive
constant $\epsilon $ and rapidly decreases outside the region $\left(
x_{B}-\epsilon ,x_{B}+\epsilon \right) $. Also, we assume that

\begin{equation*}
w_{B}\left( x\right) p_{B}\left( x\right) =p_{B}\left( x\right) .
\end{equation*}%
In order to calculate $\limfunc{Tr}\left[ \rho _{F}\hat{H}_{B}\right] $,
recall that the Schr\H{o}dinger operators in $\hat{\varepsilon}\left(
x\right) $ are transformed by $U_{n}\left( B\right) $ via 
\begin{equation*}
U_{n}^{\dag }\left( B\right) \hat{\phi}\left( x\right) U_{n}\left( B\right) =%
\hat{\phi}\left( x\right) ,
\end{equation*}%
\begin{equation*}
U_{n}^{\dag }\left( B\right) \hat{\Pi}\left( x\right) U_{n}\left( B\right) =%
\hat{\Pi}\left( x\right) +ga_{n}p_{B}\left( x\right) .
\end{equation*}%
Using these relationships, we obtain%
\begin{equation*}
U_{n}^{\dag }\left( B\right) \hat{H}_{B}U_{n}\left( B\right) =\hat{H}%
_{B}+ga_{n}\hat{O}_{B}+\frac{g^{2}}{2}a_{n}^{2}\int p_{B}\left( x\right)
^{2}dx,
\end{equation*}%
where operator $\hat{O}_{B}$ is defined by%
\begin{equation*}
\hat{O}_{B}=\int p_{B}\left( x\right) \hat{\Pi}(x)dx.
\end{equation*}%
The localized energy $\langle \hat{H}_{B}\rangle =\limfunc{Tr}\left[ \rho
_{F}\hat{H}_{B}\right] $ is then given by%
\begin{equation*}
\langle \hat{H}_{B}\rangle =\sum_{n}\langle 0|M_{n}^{\dag }\left( A\right)
\left( U^{\dag }(T)U_{n}^{\dag }\left( B\right) \hat{H}_{B}U_{n}\left(
B\right) U(T)\right) M_{n}\left( A\right) |0\rangle ,
\end{equation*}%
where

\begin{eqnarray}
&&U^{\dag }(T)U_{n}^{\dag }\left( B\right) \hat{H}_{B}U_{n}\left( B\right)
U(T)  \notag \\
&=&U^{\dag }(T)\hat{H}_{B}U(T)+ga_{n}\hat{O}_{B}(T)+\frac{g^{2}}{2}%
a_{n}^{2}\int p_{B}\left( x\right) ^{2}dx,  \label{208}
\end{eqnarray}%
and $\hat{O}_{B}(T)=U^{\dag }(T)\hat{O}_{B}U(T)$. It is noted that the above
operator commutes with $M_{n}\left( A\right) $ at time $T$. This is because
the relations 
\begin{eqnarray}
\left[ U^{\dag }(T)\hat{H}_{B}U(T),~M_{n}(A)\right] &=&0,  \label{201} \\
\left[ ga_{n}\hat{O}_{B}(T),~M_{n}(A)\right] &=&0  \label{202}
\end{eqnarray}%
hold. Equations (\ref{201}) and (\ref{202}) can be verified through Eq.(\ref%
{104}) and a relation obtained by differentiation of Eq.(\ref{103}) such
that 
\begin{align*}
\partial _{x}\hat{\phi}(t,x)& =\frac{1}{2}\left[ \partial _{x}\hat{\phi}%
(x+t)+\partial _{x}\hat{\phi}(x-t)\right] \\
& +\frac{1}{2}\left[ \hat{\Pi}(x+t)-\hat{\Pi}(x-t)\right] .
\end{align*}%
Thus, we are able to obtain a relationship such that 
\begin{equation}
\langle \hat{H}_{B}\rangle =\sum_{n}\langle 0|M_{n}^{\dag }\left( A\right)
M_{n}\left( A\right) \left( U^{\dag }(T)U_{n}^{\dag }\left( B\right) \hat{H}%
_{B}U_{n}\left( B\right) U(T)\right) |0\rangle .  \label{108}
\end{equation}%
Substituting Eq.(\ref{208}) into Eq.(\ref{108}) yields the following:%
\begin{align*}
\langle \hat{H}_{B}\rangle & =\langle 0|\left( \sum_{n}M_{n}^{\dag }\left(
A\right) M_{n}\left( A\right) \right) U^{\dag }(T)\hat{H}_{B}U(T)|0\rangle \\
& +g\langle 0|\left( \sum_{n}a_{n}M_{n}^{\dag }\left( A\right) M_{n}\left(
A\right) \right) \hat{O}_{B}(T)|0\rangle \\
& +\frac{g^{2}}{2}\int p_{B}\left( x\right) ^{2}dx\langle 0|\left(
\sum_{n}a_{n}^{2}M_{n}^{\dag }\left( A\right) M_{n}\left( A\right) \right)
|0\rangle .
\end{align*}%
Let us define Hermitian operators $\hat{D}_{A}$ and $\tilde{D}_{A}^{2}$ as%
\begin{equation}
\hat{D}_{A}=\sum_{n}a_{n}M_{n}^{\dag }\left( A\right) M_{n}\left( A\right) ,
\label{101}
\end{equation}%
\begin{equation}
\tilde{D}_{A}^{2}=\sum_{n}a_{n}^{2}M_{n}^{\dag }\left( A\right) M_{n}\left(
A\right) .  \label{102}
\end{equation}%
Using Eqs.(\ref{5}), (\ref{101}) and (\ref{102}), $\langle \hat{H}%
_{B}\rangle $ can be simplified into: 
\begin{align*}
\langle \hat{H}_{B}\rangle & =\langle 0|U^{\dag }(T)\hat{H}_{B}U(T)|0\rangle
\\
& +g\langle 0|\hat{D}_{A}\hat{O}_{B}(T)|0\rangle \\
& +\frac{g^{2}}{2}\int p_{B}\left( x\right) ^{2}dx\langle 0|\tilde{D}%
_{A}^{2}|0\rangle .
\end{align*}%
Because $U(T)|0\rangle =|0\rangle $ and $\langle 0|\hat{H}_{B}|0\rangle =0$,
the first term on the right-hand side vanishes. Hence, $\langle \hat{H}%
_{B}\rangle $ is given by

\begin{equation*}
\langle \hat{H}_{B}\rangle =\frac{1}{2}\xi g^{2}+\eta g,
\end{equation*}%
where constants $\xi $ and $\eta $ are defined as

\begin{equation}
\xi =\langle 0|\tilde{D}_{A}^{2}|0\rangle \int p_{B}(x)^{2}dx,  \label{301}
\end{equation}%
\begin{equation}
\eta =\langle 0|\hat{D}_{A}\hat{O}_{B}(T)|0\rangle .  \label{302}
\end{equation}%
By fixing the parameter $g$ by%
\begin{equation*}
g=-\frac{\eta }{\xi },
\end{equation*}%
we obtain a negative value for $\langle \hat{H}_{B}\rangle $ given by%
\begin{equation}
\langle \hat{H}_{B}\rangle =-\frac{\eta ^{2}}{2\xi }<0.  \label{303}
\end{equation}%
It can be shown explicitly from Eq.(\ref{7}) that expectation value of
energy density is exactly zero right before the operation in phase (4). This
is because $\rho (T)$ in Eq. (\ref{7}) is a local vacuum state (or strictly
localized state) in which physical properties around $B$ are the same of
those of vacuum. Vanishing of $\limfunc{Tr}\left[ \rho (T)\hat{H}_{B}\right] 
$ is regarded as an example of general results about statistical
independence of separable localized regions (see e.g. \cite{BS}). Because $%
\langle \hat{H}_{B}\rangle $ becomes negative shortly after phase (4), this
field system releases positive energy to Bob's apparatus $U_{n}\left(
B\right) $. The amount of energy is given by $E_{B}=-\langle \hat{H}%
_{B}\rangle =\eta ^{2}/(2\xi )\,$. We note that for the quantum state $\rho
(T)$ in Eq.(\ref{7}) many-point functions of the field are equal to those of
the vacuum state in the vicinity of Bob. Consequently, we can regard phase
(4) as an energy-extraction process from the local vacuum of $\phi $ around
Bob. It is a typical property of negative energy that negative-energy
wavepackets generated by Bob cannot evolve independently of positive-energy
wavepackets generated first by Alice. As mentioned in section 3, this is
because the existence of negative energy is sustained by the existence of
positive energy so as to make the total energy in space nonnegative.
Negative energy density is able to emerge only in spatial regions that have
finite correlation to other spatial regions with positive energy density.
Thus, it is impossible to separate a wavepacket with a fixed negative energy
far from wavepackets with positive energy. This observation teaches us that
Bob's wavepackets begin to chase after Alice's wavepackets and form a
loosely bound state with them after phase (4). At first glance, this
statement about loosely bound states might seem irrelevant because the
system is in one spatial dimension and both Alice's and Bob's wavepackets
maintain their interval while propagating with the same velocity. However,
in higher-dimensional field theory, the traveling direction of the first
positive-energy wavepackets generally do not have an isotropic distribution
and, in particular, there may be a spatial region through which no
wavepacket passes. If Bob stays at a point in such a region and makes his
local operations, formation of the above loosely bound states becomes a
rather nontrivial phenomenon.

\bigskip

\section{Example}

\bigskip

Although the protocol works for any general measurement by Alice, we present
a simple example of a two-valued general measurement. This will allow us to
experiment with a similar protocol extended for the electromagnetic field.
Let us choose $M_{n}(A)$ as follows:

\begin{align}
M_{0}(A)& =\cos \hat{\Phi}_{A},  \label{13} \\
M_{1}(A)& =\sin \hat{\Phi}_{A},  \label{14}
\end{align}%
where $\hat{\Phi}_{A}$ is a Hermitian operator given by

\begin{equation*}
\hat{\Phi}_{A}=\frac{\pi }{4}-\int \lambda _{A}(x)\hat{\Pi}(x)dx,
\end{equation*}%
and $\lambda _{A}(x)$ is a real localized function around Alice's location.
The POVM can be constructed by combining the system with a two-state probe
system $P$ by a certain interaction. Let us consider an orthonormal state
basis $\left\{ |0_{P}\rangle ,|1_{P}\rangle \right\} $ of $P$. Then let us
give an interaction Hamiltonian defined by

\begin{equation*}
H_{p}=ig(t)\hat{\Phi}_{A}\otimes \left[ |1_{P}\rangle \langle
0_{P}|-|0_{P}\rangle \langle 1_{P}|\right] .
\end{equation*}%
Using the time evolution operator $V(t)=\exp \left[ -i\int_{0}^{t}g(t)dtH_{p}%
\right] $, it can be easily proven that the measurement is reproduced at
time satisfying $\int_{0}^{t}g(t)dt=1$ by an ideal measurement of an
observable $|0_{P}\rangle \langle 0_{P}|-|1_{P}\rangle \langle 1_{P}|$ for
the probe system:

\begin{equation*}
M_{b}\rho M_{b}^{\dag }=\limfunc{Tr}_{P}\left[ \left( I\otimes |b_{P}\rangle
\langle b_{P}|\right) V\left( \rho \otimes |0_{P}\rangle \langle
0_{P}|\right) V^{\dag }\right] ,
\end{equation*}%
where $b=0,1$ and $\rho $ is an arbitrary state of the system. In the above
protocol setting, it is assumed that switching of the\ measurement
interaction is performed abruptly such that

\begin{equation*}
g(t)=\delta (t-0).
\end{equation*}%
To make our argument more concrete, let us choose parameters $a_{n}$ as $%
a_{n}=(-1)^{n}$. From Eqs.(\ref{301}) and (\ref{302}), the following
explicit relations are derived:

\begin{equation}
\hat{D}_{A}=\sin \left( 2\int \lambda _{A}(x)\hat{\Pi}(x)dx\right) ,
\label{304}
\end{equation}

\begin{equation}
\tilde{D}_{A}^{2}=I.  \label{305}
\end{equation}%
From Eq.(\ref{10}), the measurement by Alice yields post-measurement states
depending on $n$ as a sum of two coherent states given by

\begin{equation*}
|A_{0}\rangle =\frac{1}{\sqrt{2}}\left[ e^{i\frac{\pi }{4}}|\lambda
_{A}\rangle +e^{-i\frac{\pi }{4}}|-\lambda _{A}\rangle \right] ,
\end{equation*}%
\begin{equation*}
|A_{1}\rangle =\frac{1}{\sqrt{2}}\left[ e^{-i\frac{\pi }{4}}|\lambda
_{A}\rangle +e^{i\frac{\pi }{4}}|-\lambda _{A}\rangle \right] ,
\end{equation*}%
where $|\lambda \rangle $ is a coherent state satisfying $\langle \lambda |%
\hat{\phi}\left( x\right) |\lambda \rangle =\lambda (x)$ and $\langle
\lambda |\hat{\Pi}(x)|\lambda \rangle =0$. For both post-measurement states,
the expectational value of the Heisenberg energy density operator takes the
value given by

\begin{equation}
\langle A_{n}|\hat{\varepsilon}\left( t,x\right) |A_{n}\rangle =\frac{1}{2}%
\left[ \left( \partial _{x}\lambda _{A}\left( x-t\right) \right) ^{2}+\left(
\partial _{x}\lambda _{A}\left( x+t\right) \right) ^{2}\right] .  \label{11}
\end{equation}%
Here it should be noticed that energy density vanishes outside the compact
supports of $\lambda _{A}\left( x-t\right) $ and $\lambda _{A}\left(
x+t\right) $ in Eq.(\ref{11}) because of locality. The first term on the
r.h.s. of Eq.(\ref{11}) describes a right-moving positive-energy wavepacket
with light velocity. The second term describes a left-moving wavepacket. The
energy input $E_{A}$ is given by integration of Eq.(\ref{11}) as

\begin{equation}
E_{A}=\int_{-\infty }^{\infty }\left( \partial _{x}\lambda _{A}\left(
x\right) \right) ^{2}dx.  \label{Ea}
\end{equation}

At time $t=T$, Bob gets information about $n$ and performs $U_{n}\left(
B\right) $ to the state. \ The amount of energy gain by Bob can be
calculated from Eq.(\ref{303}) as follows. First, $\xi $ is obtained from
Eq.(\ref{301}) by%
\begin{equation}
\xi =\int p_{B}(x)^{2}dx.  \label{xi}
\end{equation}%
On the basis of Eq.(\ref{302}) and a relation such that

\begin{align*}
\hat{O}_{B}(T)& =\frac{1}{2}\int dx\left[ p_{B}\left( x-T\right)
+p_{B}\left( x+T\right) \right] \hat{\Pi}(x) \\
& +\frac{1}{2}\int dx\left[ p_{B}\left( x-T\right) -p_{B}\left( x+T\right) %
\right] \partial _{x}\hat{\phi}\left( x\right) ,
\end{align*}%
it is possible to write $\eta $ as 
\begin{align}
\eta & =\int \langle 0|\hat{D}_{A}\hat{\Pi}(x)|0\rangle \left[ p_{B}\left(
x-T\right) +p_{B}\left( x+T\right) \right] dx  \notag \\
& +\int \langle 0|\hat{D}_{A}\partial _{x}\hat{\phi}\left( x\right)
|0\rangle \left[ p_{B}\left( x-T\right) -p_{B}\left( x+T\right) \right] dx.
\label{401}
\end{align}%
It is worth noting that the following relation holds from Eq.(\ref{304}).

\begin{eqnarray*}
\langle 0|\hat{D}_{A} &=&\frac{1}{2i}\langle 0|\left[ \exp \left( 2i\int
\lambda _{A}(x)\hat{\Pi}(x)dx\right) -\exp \left( -2i\int \lambda _{A}(x)%
\hat{\Pi}(x)dx\right) \right] \\
&=&\frac{1}{2i}\left[ \langle -2\lambda _{A}|-\langle 2\lambda _{A}|\right] .
\end{eqnarray*}%
and that%
\begin{equation*}
\langle 2\lambda _{A}|0\rangle =\langle -2\lambda _{A}|0\rangle =\langle
2\lambda _{A}|0\rangle ^{\ast }.
\end{equation*}%
Let us introduce a distributional function $\Delta (x)$ by

\begin{equation*}
\Delta (x)=2\langle 0|\dot{\phi}(0,x)\dot{\phi}(0,0)|0\rangle =\frac{1}{2\pi 
}\int_{-\infty }^{\infty }|k|e^{ikx}dk.
\end{equation*}%
$\Delta (x)$ has a delta-functional contribution at $x=0$ and is evaluated as

\begin{equation*}
\Delta (x)=-\frac{1}{\pi \left\vert x\right\vert ^{2}}
\end{equation*}%
for $x\neq 0$. By using $\Delta (x)$, we derive the following relationship:

\begin{equation*}
\langle 2\lambda _{A}|\hat{\Pi}\left( x\right) |0\rangle =i\langle 2\lambda
_{A}|0\rangle \int_{-\infty }^{\infty }\Delta (x-y)\lambda _{A}(y)dy.
\end{equation*}%
Hence, we obtain the relation:

\begin{equation}
\langle 0|\hat{D}_{A}\hat{\Pi}(x)|0\rangle =-\langle 2\lambda _{A}|0\rangle
\int_{-\infty }^{\infty }\Delta (x-y)\lambda _{A}(y)dy.  \label{501}
\end{equation}%
We are also able to show that

\begin{eqnarray}
&&\int \langle 0|\hat{D}_{A}\partial _{x}\hat{\phi}\left( x\right) |0\rangle %
\left[ p_{B}\left( x-T\right) -p_{B}\left( x+T\right) \right] dx  \notag \\
&=&i\langle 2\lambda _{A}|0\rangle \int \partial _{x}\lambda _{A}(x)\left[
p_{B}\left( x-T\right) -p_{B}\left( x+T\right) \right] dx=0.  \label{502}
\end{eqnarray}%
Here, the last integral vanishes because there is no overlap between $%
\partial _{x}\lambda _{A}(x)$ and $p_{B}\left( x\pm T\right) $. Substituting
Eqs.(\ref{501}) and (\ref{502}) into Eq.(\ref{401}) yields

\begin{equation}
\eta =-\langle 2\lambda _{A}|0\rangle \int dx\int dy\lambda _{A}(x)\left[
\Delta (x-y-T)+\Delta (x-y+T)\right] p_{B}\left( y\right) .  \label{eta1}
\end{equation}%
By substituting Eqs.(\ref{xi}) and (\ref{eta1}) into Eq.(\ref{303}), we
obtain the final expression for $E_{B}$:

\begin{equation}
E_{B}=\frac{\left( \langle 2\lambda _{A}|0\rangle \int dx\int dy\lambda
_{A}(x)\left[ \frac{1}{\left( x-y-T\right) ^{2}}+\frac{1}{\left(
x-y+T\right) ^{2}}\right] p_{B}\left( y\right) \right) ^{2}}{2\pi ^{2}\int
p_{B}(z)^{2}dz}.  \label{res}
\end{equation}

\bigskip

Extension of the protocol to the 3+1 dimensional electromagnetic field is
also possible by adopting three-dimensional measurements and unitary
operations. The free gauge field $A^{\mu }$ can be expressed in the Coulomb
gauge. The gauge fixing condition is given by $A^{0}=0$ and $\func{div}\vec{A%
}=0$. The two-valued general measurement operators corresponding to Eqs.(\ref%
{13}) and (\ref{14}) are defined by setting 
\begin{equation*}
\hat{\Phi}_{A}=\frac{\pi }{4}-\int \vec{\lambda}_{A}(\vec{x})\cdot \vec{E}(%
\vec{x})d^{3}x,
\end{equation*}%
where $\vec{E}(\vec{x})$ is the electric field and $\vec{\lambda}_{A}(\vec{x}%
)$ is a three-dimensional vector local function around Alice's position. The
local unitary operation of Bob in Eq.(\ref{16}) is extended as

\begin{equation*}
U_{n}\left( B\right) =\exp \left[ iga_{n}\int \vec{p}_{B}(\vec{x})\cdot \vec{%
A}(\vec{x})d^{3}x\right] ,
\end{equation*}%
where $\vec{p}_{B}(\vec{x})$ is a three-dimensional vector localized
function around Bob's position and should satisfy the relation

\begin{equation*}
\func{div}\vec{p}_{B}=0,
\end{equation*}%
because of residual gauge symmetry of the gauge fixing. \ A detailed
analysis on the electromagnetic field case will be published elsewhere.
Experimental checks of the protocol proposed in this paper may be promising
in quantum optics, and stimulate future development of new methods of
quantum energy transportation.

\bigskip

\section{Conclusion}

\bigskip

This paper discusses local vacuum states excited by a local general\
measurement for the vacuum state of a free massless scalar field in 1+1
dimensions. Properties of local vacuum states are the same as those of
vacuum as long as we consider the vanishing-energy regions of those states.
A protocol is presented that can partially extract the excitation energy
from local vacuum states using both information on the measurement result
and a quantum apparatus located away from the measurement point, even if the
field has, on average, no quantum energy around the apparatus. As an
example, the case of the two-valued general measurements defined in Eqs.(\ref%
{13}) and (\ref{14}) are analyzed in detail. The energy input for the
measurement is given by the result in Eq.(\ref{Ea}). The extracted energy is
calculated using Eq.(\ref{res}) for measurement-data-dependent unitary
operations given by Eq.(\ref{16}) with $a_{n}=(-1)^{n}$ for $n=0,1$.

\bigskip

\textbf{Acknowledgments}\newline

\bigskip

I would like to thank I.Ojima for giving me useful information. This
research was partially supported by the SCOPE project of the MIC.

\bigskip

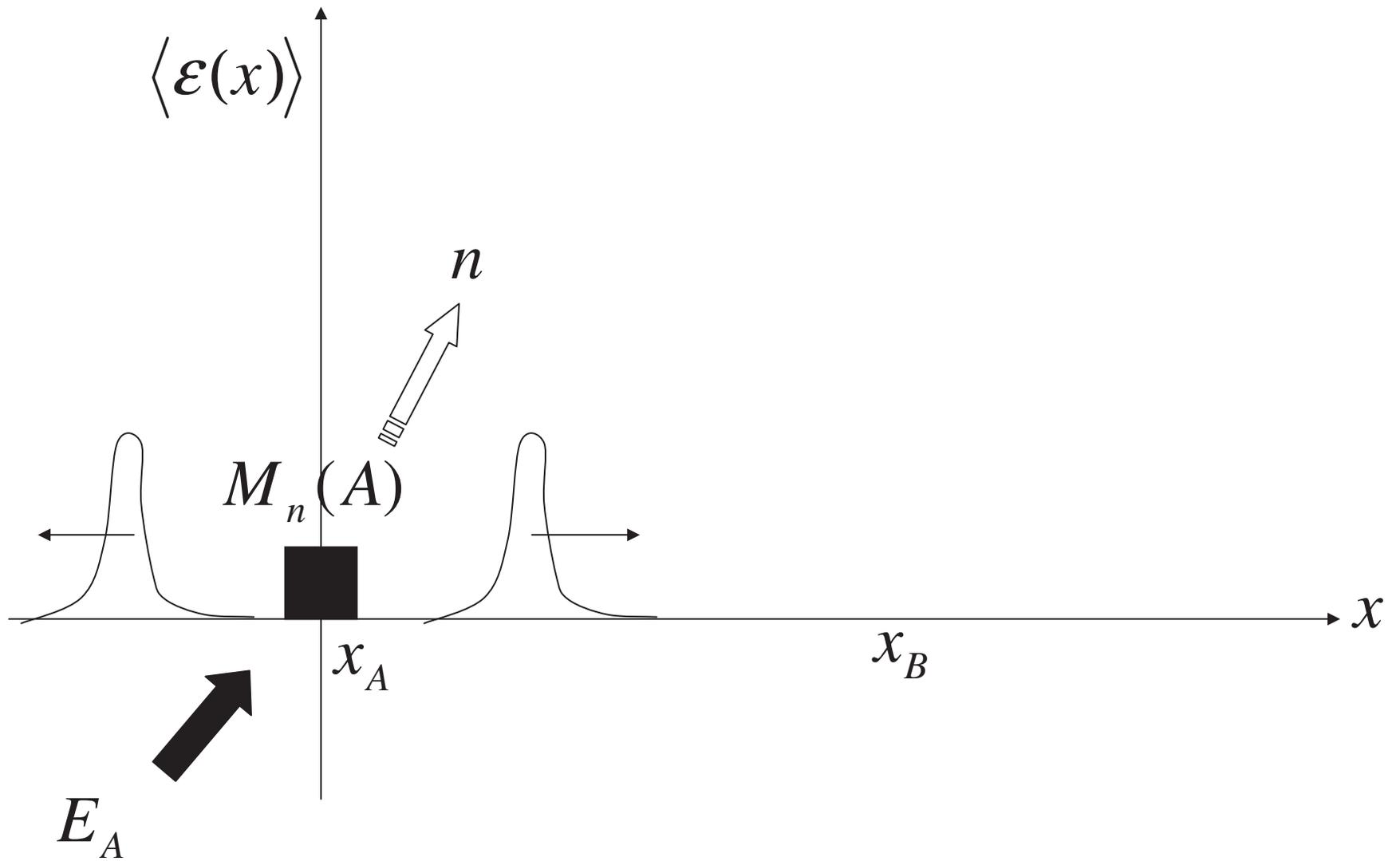
Figure 1: The first schematic diagram of QFET. Alice stays at $x=x_{A}$ and
Bob at $x=x_{B}$. Alice performs a general measurement to the vacuum state
with energy input $E_{A}~$and obtains the measurement result $n$. Then,
positive-energy wavepackets are generated in the system and escape to
spatial infinity at the speed of light. The expectational value of the
energy density $\langle \varepsilon (x)\rangle =\limfunc{Tr}\left[ \rho \hat{%
\varepsilon}(x)\right] $ plotted as a function of $x$.

\bigskip

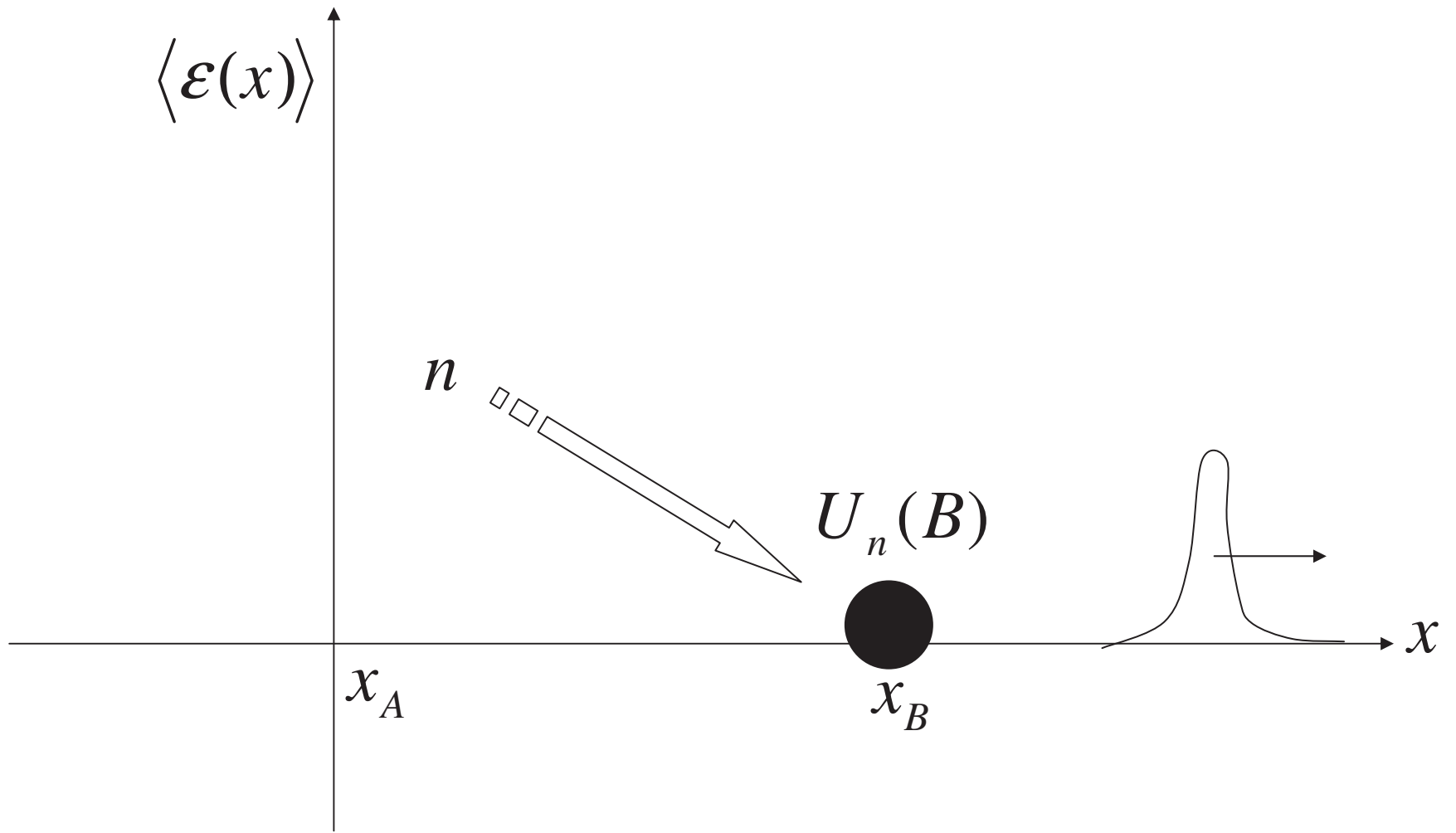
Figure 2: The second schematic diagram of QFET. After the wavepacket passes
through Bob's location, Alice announces to Bob the measurement result $n$.
Bob obtaining $n$ performs local unitary operation $U_{n}(B)$ depending on $%
n $.

\bigskip

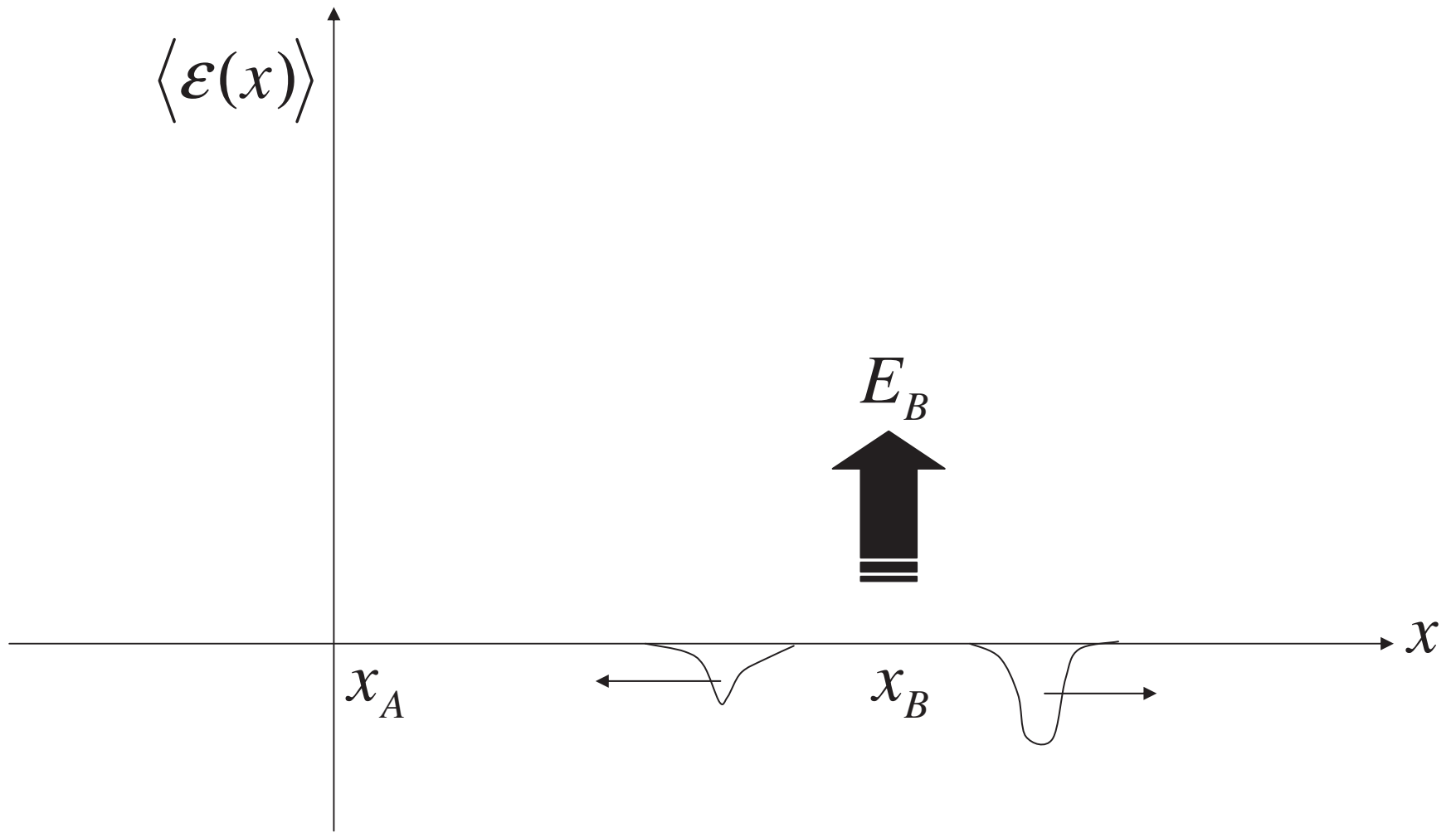
Figure 3: The third schematic diagram of QFET. In the process of $U_{n}(B)$,
Bob gets a positive amount of energy from the field, generating
negative-energy wavepackets in the field system.

\bigskip

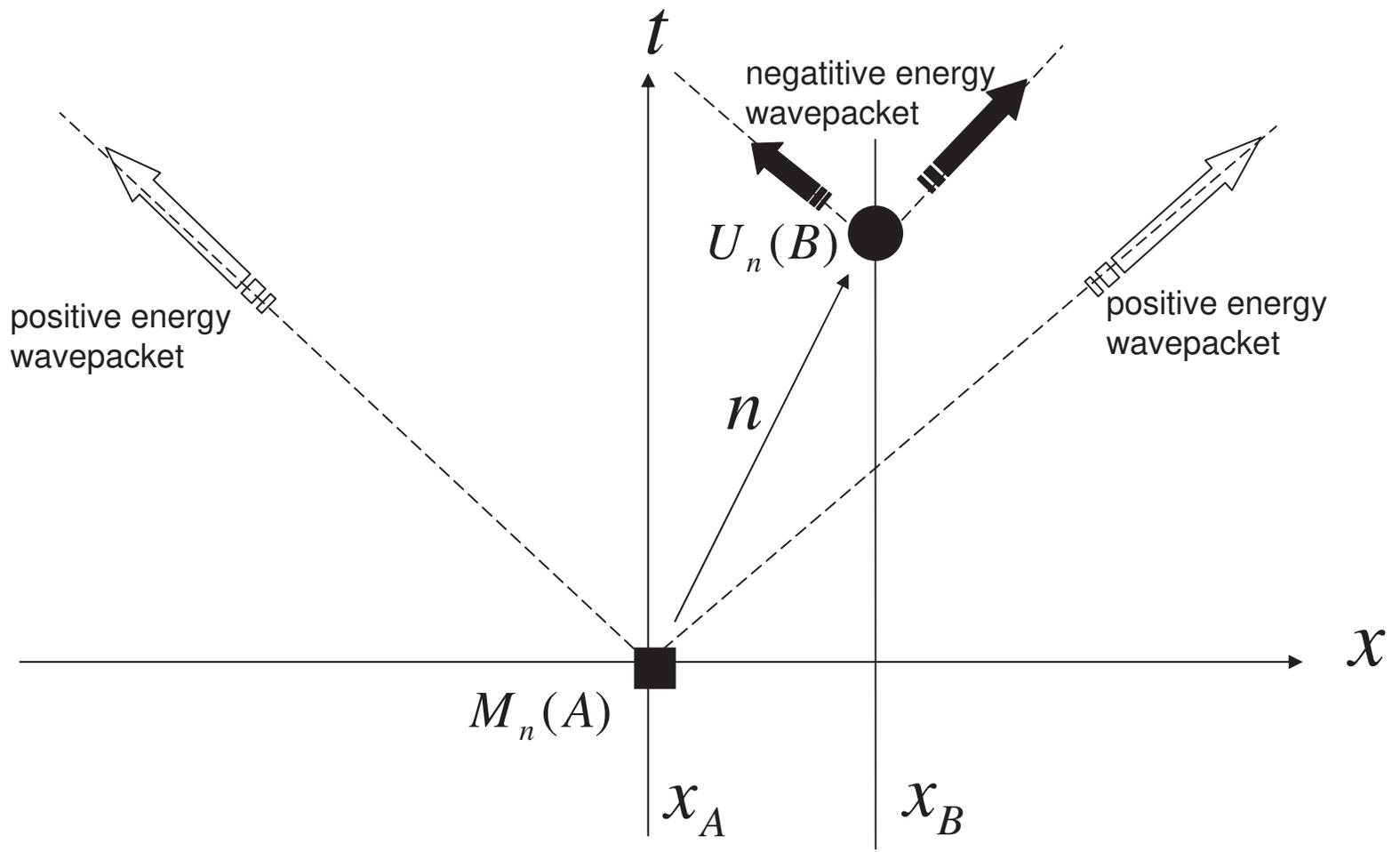
Figure 4: A spacetime diagram of QFET. Entanglement is created between
positive-energy wavepackets generated by Alice and negative-energy
wavepackets by Bob, which form a loosely bound state.

\bigskip

\end{document}